\documentclass{article}
\usepackage{siunitx}

\usepackage{authblk}
\usepackage{amssymb}
\usepackage{amsmath}
\usepackage{gensymb}
\usepackage{graphicx}
\usepackage{graphics}
\usepackage{geometry}
\usepackage{lipsum}
\usepackage{times}
\usepackage{xcolor}
\usepackage{xspace}
\usepackage{textcomp}
\usepackage{bm}
\usepackage{relsize}
\usepackage{float}
\usepackage[mathscr]{euscript}
\usepackage[font=footnotesize,labelfont=bf]{caption}
\usepackage[font=footnotesize,labelfont=bf]{subcaption}
\usepackage{multirow}
\usepackage[utf8]{inputenc}

\newcommand{\fref}[1] {Fig.~\ref{#1}\xspace}
\newcommand{\tref}[1] {Table~\ref{#1}\xspace}
\newcommand{\argmin}[1] {\underset{#1}{\textnormal{arg min}}}
\newcommand{\vect}[1]{\mathbf{\underline{#1}}}
\newcommand{\xmath}[1] {\ensuremath{#1}\xspace}
\newcommand{\vtheta} {\xmath{\vect{\theta}}}

\newcommand{\F} {\xmath{\bm{F}}}
\newcommand{\W} {\xmath{\bm{W}}}

\newcommand{\mtis}{M_{\text{tis}}}
\newcommand{\mart}{M_{\text{art}}}

\title{Optimizing MRF-ASL Scan Design for Precise Quantification
of Brain Hemodynamics using Neural Network Regression}
\author[1]{Anish Lahiri}
\author[1]{Jeffrey A Fessler}
\author[2]{Luis Hernandez-Garcia}
\affil[1]{Department of Electrical and Computer Engineering, University of Michigan}
\affil[2]{Functional MRI Laboratory, University of Michigan}
\date{}

\begin{document}

\maketitle

\textbf{Corresponding Author:} Anish Lahiri

\textbf{Address:} Functional MRI Laboratory, 2360 Bonisteel Blvd, Ann Arbor, MI 48105, USA. 

\textbf{E-mail:} anishl@umich.edu



\newpage
\begin{abstract}
    \textbf{Purpose: }Arterial Spin Labeling (ASL) is a quantitative, non-invasive alternative to perfusion imaging with contrast agents. Fixing values of certain model parameters in traditional ASL, which actually vary from region to region, may introduce bias in perfusion estimates. Adopting Magnetic Resonance Fingerprinting (MRF) for ASL is an alternative where these parameters are estimated alongside perfusion, but multiparametric estimation can degrade precision. We aim to improve the sensitivity of ASL-MRF signals to underlying parameters to counter this problem, and provide precise estimates. We also propose a regression based estimation framework for MRF-ASL.

    \textbf{Methods: }To improve the sensitivity of MRF-ASL signals to underlying parameters, we optimize ASL labeling durations using the Cramer-Rao Lower Bound (CRLB). This paper also proposes a neural network regression based estimation framework trained using noisy synthetic signals generated from our ASL signal model. 
    
    \textbf{Results: }We test our methods in silico and in vivo, and compare with multiple post labeling delay (multi-PLD) ASL and unoptimized MRF-ASL. We present comparisons of estimated maps for six parameters accounted for in our signal model.
    
    \textbf{Conclusions: }The scan design process facilitates precise estimates of multiple hemodynamic parameters and tissue properties from a single scan, in regions of gray and white matter, as well as regions with anomalous perfusion activity in the brain. The regression based estimation approach provides perfusion estimates rapidly, and bypasses problems with quantization error.
    
\end{abstract}

\section{Introduction}

Quantitative imaging of tissue properties is gaining increasing prominence in the diagnosis, prognosis and treatment planning of several diseases, e.g.,  \cite{Gatenby2013QuantitativeEcology,Lee2008QuantitativeMasses,Rosenthal1992QuantitativeDisease.,Ramani2006QuantitativeDisease}. Being able to move away from the uncertainties or variations usually associated with qualitative intensity-based imaging allows gleaning information focused on the physiological phenomena being investigated. This approach can benefit all stages of the imaging pipeline in clinical practice. In the context of cerebrovascular disorders in particular, quantitative perfusion imaging has found several applications \cite{Detre2012ApplicationsBrain,Kucharczyk1991Diffusion/perfusionIschemia,Watts2013ClinicalLabeling,Deibler2008ArterialArtifacts.,Ho2008Radiologic-pathologicHemangioblastoma.,Boxerman2006RelativeNot.,Haller2016ArterialApplications}, in the brain and beyond \cite{DeBazelaire2005ArterialCarcinoma1.,Bolar2006QuantificationASL-FAIRER}. Typically, quantitative imaging of perfusion involves gadolinium-based contrast enhanced MRI, limiting its use in clinical applications due to the lack of fast repeatability and risks involved in cases of subjects with nephrogenic disorders \cite{Wang2011IncidenceGuidelines,Roditi2009RenovascularEra}. 

Arterial Spin Labeling \cite{Detre1992PerfusionImaging} provides an alternative to contrast agent based MRI by magnetically labeling blood flowing into organs or tissues of interest. This approach involves temporary inversion of the spins present in flowing blood upstream of the organ under scrutiny, by applying radiofrequency (RF) magnetic pulses. These inverted spins then behave like an \textit{endogenous} tracer that can be detected when it perfuses into the tissue in the relevant organ shortly afterwards. This method is non-invasive, non-toxic, quickly repeatable and has a much simpler workflow than contrast enhanced MRI. However, ASL images are limited by low spatial and temporal Signal-to-Noise Ratio (SNR) \cite{Perthen2008SNRT,Wong1998AImaging,Chen2011Test-retestStrategies}, because of the need for the subtraction of label and control images. This drawback is more pronounced in white matter, where the performance of traditional ASL methods is poor\cite{vanGelderen2008PittfallsLabeling}. Estimating perfusion using this method requires knowledge of a number of tissue properties or hemodynamic factors which are usually fixed to literature values, whereas in reality, some of these (tissue $T_1$ for example), vary significantly from region to region. Fixing these parameter values lead to significant biases in perfusion estimates while efforts at estimating such factors from separate scans can be undesirably time-consuming.

Magnetic Resonance Fingerprinting (MRF) is a recently developed technique \cite{Ma2013MagneticFingerprinting} where multiple hemodynamic parameters and tissue properties can be estimated simultaneously from a single acquisition, thereby improving accuracy at the possible expense of precision in estimates. Nevertheless, information accrued from the additional estimated parameter maps may aid understanding of physiologial conditions. MRF utilizes transient state signals obtained by varying imaging parameters such as the repetition time (TR) or flip angles as identifiers for the underlying physiological factors. A standard approach to multiparametric estimation using such a technique involves searching through `dictionaries' consisting of signals generated by feasible combinations of parameters, in a Maximum Likelihood manner. In an ASL based fingerprinting \cite{Su2017MultiparametricASL,Wright2018EstimationLabeling} setting, the observed signal depends on several parameters (typically 5-7), presenting a considerable challenge to precise estimation. For example, with more parameters to estimate, it becomes difficult to maintain and search a `fine' dictionary. Specifically, \cite{Su2017MultiparametricASL} reports a $2$ hr estimation time for a single slice, with a dictionary quantization of 6mL/100g/min for perfusion. 

As an alternative, this paper uses a regression based estimator to generate predictions from fingerprint data. While the use of regressors for MR Fingerprinting based estimation has become more prevalent recently \cite{Nataraj2018Dictionary-FreeKernels,Cohen2018MRDRONE}, our previous work \cite{Lahiri2018OptimizedRegression} was the first to investigate neural network regression for ASL Fingerprinting, where there are considerably more parameters to estimate. Estimation using neural network regression allows for much faster estimation, and overcomes quantization error. 

Regardless of the estimation technique used, if the ASL fingerprints themselves are insufficiently sensitive to the underlying parameters, then estimates obtained from them will lack precision. Thus, the first goal of this paper is to increase the information conveyed by fingerprints, using Cramer-Rao bound based optimization of scan parameters in ASL. An example of such scan parameters are the labeling durations in the scan. While there has been some work on optimizing scan-design for MR sequences in quantitative imaging \cite{Nataraj2016OptimizingSequences,Zhao2017OptimalDynamics}, and even specifically in ASL \cite{Xie2008OptimalExperiments,Woods2019AExperiments}, our work is the first to investigate it in an ASL fingerprint setting. While most other pertinent focused on providing precise results in regions of gray matter, we use a cost function having a comprehensive uniform prior, enabling precise estimation over a wide range of feasible parameter values, including white matter or potential anomalies. We also constrain our optimization procedure to adhere to a fixed scan time for practicality. The primary focus of our work is to establish the need for scan design optimization regardless of the estimation technique involved. Through our work, we establish that optimized scan design coupled with regression-based estimation should further the transition of ASL Fingerprinting to clinical use. 

The rest of the paper is organized as follows: Section \ref{Met} introduces the ASL signal model used in this work. This model, along with a Cramer-Rao bound based cost function, is then used to optimize our scan design. Next, we design a neural network regressor for estimating hemodynamic parameters and tissue properties using  fingerprints simulated with a combination of the optimized scan design and the described model. We also devise a post-processing technique to mitigate nuisance effects in our acquisitions. Thereafter, we describe the creation of in-silico datasets to test the performance of our methods, as well the methods we compare to in our work. We then describe the in-vivo experiments we performed in the validation of our designed methods. Section \ref{Res} shows the theoretical predictions of the performance of our scan design, as well as the results of comparisons in-silico and in-vivo with other methods, namely two suboptimal MRF ASL scans and multi-PLD ASL. Section \ref{Dis} elaborates upon these results and the inferences we draw from them. Section \ref{Conc} describes our conclusions. 

\section{Methods}\label{Met}

\subsection{ASL Signal Model}\label{modT}
\begin{figure}[H]
\begin{center}
\includegraphics[height=1.22in]{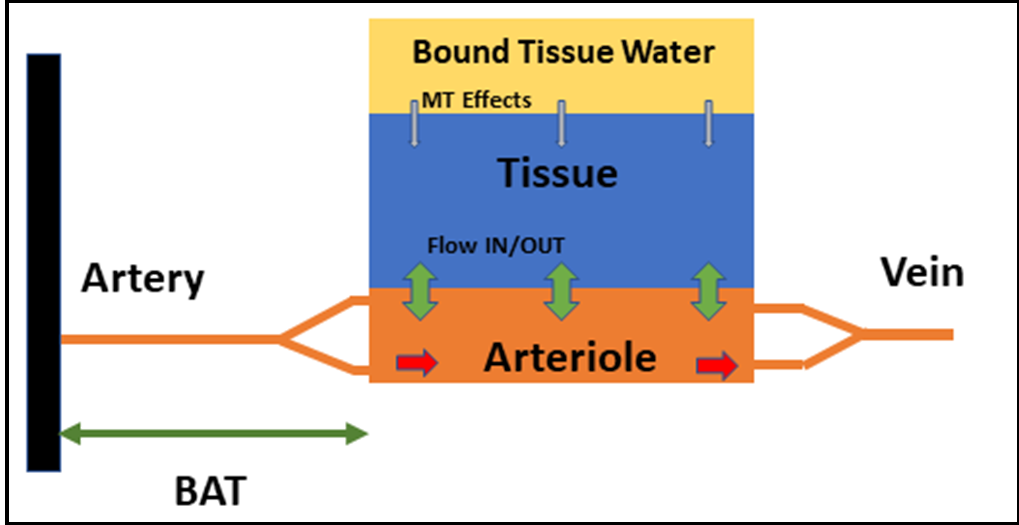}
\caption{Two compartment ASL signal model used for both optimization and estimation. The unknowns in the model are: perfusion from the arteriole to the tissue, arrival time of the labeled blood bolus at the arteriole, the magnetization transfer rate, the arterial blood volume fraction, and the relaxation time of water in tissue.}
\label{fig:model}
\end{center}
\vspace{-0.1in}
\end{figure}

To describe the ASL signal in the brain for scan design optimization and parameter estimation, we used the two-compartment model depicted in \fref{fig:model}. Although the single compartment model introduced in \cite{Buxton1998ALabeling.} has been the \textit{de-facto} standard in ASL literature in the past, several works \cite{Parkes2005QuantificationModels,Zhou2001Two-CompartmentTagging,Brookes2007NoninvasiveLabeling,Capron2013Cine-ASL:Optimization,Noguchi2012ArterialTechnique} have raised issues of  oversimplification associated with single-compartment modeling, and have adopted two-compartment models. For ASL fingerprinting, such models have been highlighted in \cite{Wright2018EstimationLabeling,Su2016MultiparametricASL}. Our chosen model for the ASL signal consists of separate compartments for blood in tissue and arteries, as well as an additional compartment to incorporate Magnetization Transfer effects. In the model, magnetically labelled blood flows into the arterial compartment through the arterioles, and perfuses into the tissue compartment therein. The Cerebral Blood Volume fraction (CBVa) determines the portion of the acquired signal to which each compartment contributes, and $T_1$ relaxation of blood and tissue is accounted for in the signal description. The longitudinal magnetization of the tissue compartment thus evolves as:
\begin{align}\label{modeleq}
   \frac{d\mtis(t)}{dt} = -\frac{\mtis^0-\mtis(t)}{T_{1,\text{tis}}} + f\cdot \mart(t) - \frac{f}{\lambda}\cdot\mtis(t)-K_m(t)\cdot\mtis(t),
\end{align}
where $\mtis$ and $\mart$ represent the magnetization in the tissue-compartment and the arterial compartment respectively, $\lambda$ is the blood-brain partition coefficient, $K_m$ is the Magnetization Transfer rate (MTR), $f$ is the rate of perfusion. Here, $T_{1,\text{art}}$ is the arterial relaxation time, and $T_{1,\text{tis}}$ (truncated to $T_1$ in later sections) the relaxation time of tissue. The input to the arterial compartment is determined by a labeling  function, which is described in eqn. \eqref{labeleq}. The arterial magnetization is described using an input or labeling function as follows:
\begin{align}\label{labeleq}
    \mart(t) = 1- 2\cdot\alpha\cdot \text{inp}(t)\cdot e^{-\frac{(t-\delta)}{T_{1,\text{art}}}},
\end{align}
where $\alpha$ is the inversion efficiency, $\delta$ is the Bolus Arrival Time (BAT), and
\begin{align*}
    \text{inp}(t) = 
    \begin{cases}
    1, &\text{for labeling pulses }\\
    0, &\text{otherwise. }
    \end{cases}
\end{align*}
 The total longitudinal magnetization can thus be described as:
\begin{align}\label{modeleqfin}
    M(t) = CBV_a\cdot\mart + (1-CBV_a)\cdot\mtis,
\end{align}
where $CBV_a$ is the Cerebral Blood Volume fraction described earlier. The observed ASL-MRF signal, $M(t)\cdot\sin(\beta)$, where $\beta$ is the flip angle, is sampled at the time(s) of acquisition, which are dictated by the scan schedule. We use signals generated using this model for both optimization of scan design as well as for training the neural network estimators. For the purposes of our work, the values of $\lambda$ and $\alpha$ were set to $0.9$ and $85\%$ respectively.

\subsection{Pulse Sequence}\label{pulsT}
In ASL Fingerprinting, each repetition time (TR) in the sequence consists of a labeling period ($T_\text{tag}$), post labeling delay ($T_\text{delay}$), followed by a small period for signal acquisition ($T_\text{aq}$), (ideally instantaneous, but usually accounted for) and a period for adjustment ($T_\text{adjust}$) before the next label/control occurs. Every pulse in the sequence can either be a label, control, or `silence' (where there is no RF excitation at all). In this work, we vary the TR by changing the labeling durations, while holding all other parameters of the pulse sequence fixed ($T_\text{delay}=55$ms, $T_\text{aq} =32.4$ms, $T_\text{adjust} = 50$ms). The TRs in our sequence are varied to generate a signal that is informative of the underlying parameters. Section \ref{optT} describes how we optimize the aforementioned labeling durations by picking from a set of candidate schedules. The label-control order (also referred to as the `label order' later) is pseudo-randomized, but has approximately equal numbers of label, control and silence pulses. Regardless of the number of pulses, or the duration of individual TRs, we ensure that the total duration of the scan is fixed. This fixed total duration can be discretionary. Here, we acquire $700$ images for our fingerprint, with a total scan duration of $600$s for a single slice.

\subsection{Optimization with CRLB}\label{optT}
A major focus of our work is to investigate the benefits of scan design optimization in ASL Fingerprinting. From an information theoretic standpoint, the total information present in a signal about the underlying parameters that generated it is independent of the estimator used to quantify the parameters themselves. For example, in a regression based estimation framework, regardless of whether kernel methods or neural networks are used, if the signals (or `fingerprints') themselves are too correlated, corresponding estimates will be imprecise. In an effort to make our fingerprints more informative or sensitive to parameters like perfusion or BAT, we use the Cramer-Rao Lower Bound (CRLB) to optimize the scan design parameters (namely, the labeling durations).

The CRLB represents the minimum variance in estimates that any unbiased estimator can achieve, for a particular signal model and noise level. We focus on magnitude image data and model the noise as real additive white Gaussian noise (AWGN) with standard deviation $\sigma$ (empirically calculated to be 0.01, which is low enough to justify the assumption of Gaussian noise in regions with sufficient SNR), and given our signal model $\vect{s}(\vect{\theta};\vect{\nu})$, where $\vect{\theta}$ represents hemodynamic parameters of interest, and $\vect{\nu}$ are the scan parameters, the CRLB is expressed as the inverse of the Fisher Information matrix. Fisher Information is the amount of information conveyed by an observable random variable about the parameters that generated it. It can be considered to be a measure of sensitivity of a signal to underlying parameters and is expressed as the matrix: 
\begin{align}
\mathbf{F}({\vect{\theta}};\vect{\nu}) = \frac{1}{\sigma^2}\cdot[\nabla_{\vect{\theta}}\vect{s}]^T[\nabla_{\vect{\theta}}\vect{s}].
\end{align}
To design a `good' fingerprinting sequence, we optimize over a set of feasible scan design parameters $\vect{\nu}\in\mathcal{V}$, which in our case are the labeling durations (hence $\nu$), by minimizing our design cost function at a representative collection or set of true parameter values,
$\mathbf{\Theta}$, spread uniformly over a range. 
We pick our `optimized' labeling schedule as the one that,  among all others in the feasible set $\mathscr{V}$, minimizes the following cost function:

\begin{align}
\hat{\nu} = \argmin{\nu \in \mathscr{V}}
~~\frac{1}{|\mathbf{\Theta}|}
\underset{\vtheta \in \mathbf{\Theta}}{\mathlarger{\mathlarger{\sum}}}
\text{Tr} \bigg(\W \cdot \frac{|\F^{-1}(\vect{\theta},\nu)|^{0.5}}{\bm{N}(\vtheta)}
\cdot \W \bigg)
\label{objfun}
,\end{align}
where \W is a diagonal weighting matrix assigning priority
to each hemodynamic parameter in the cost function
and $\bm{N}(\vtheta) = (\vect{\theta}^{0.5})(\vect{\theta}^{0.5})^T$ 
is a normalization matrix that is divided element-wise into the inverse Fisher Information matrix.

We used a type of exhaustive search to minimize
the design cost function \eqref{objfun}
to ensure that our optimized scan yields precise estimates over a range of `ground truth' parameter values, which is dictated by $\mathbf{\Theta}$. Minimizing the above expression is tantamount to minimizing the average normalized standard deviation of parameter estimates, weighted appropriately, over a dictated span of ground truth parameter values. For our experiments, we assign twice the weight to perfusion precision as to all the other parameters in the cost function, which are weighted equally. The cost is evaluated numerically. For the optimized scheme to function well even at values seen in pathological conditions, we use an adequately large feasible range for $\mathbf{\Theta}$. 
Specifically, the parameter ranges we used were: 12-90 ml/100g/min for perfusion, 0.002-0.03 for CBVa, 0.36-1.7 s for BAT, 0.01-0.03 s$^{-1}$ for MTR, 0.3-3.3 s for $T_1$ and 54-112 degrees for flip angles. $|\Theta|$, the number of points at which the normalized standard deviation is evaluated in the cost, was picked to be 50 during labeling schedule optimization, and 75 during label order optimization and 250 for final evaluation of the designed scan.


\begin{figure}[h!]
\begin{center}
\includegraphics[height=2.5in]{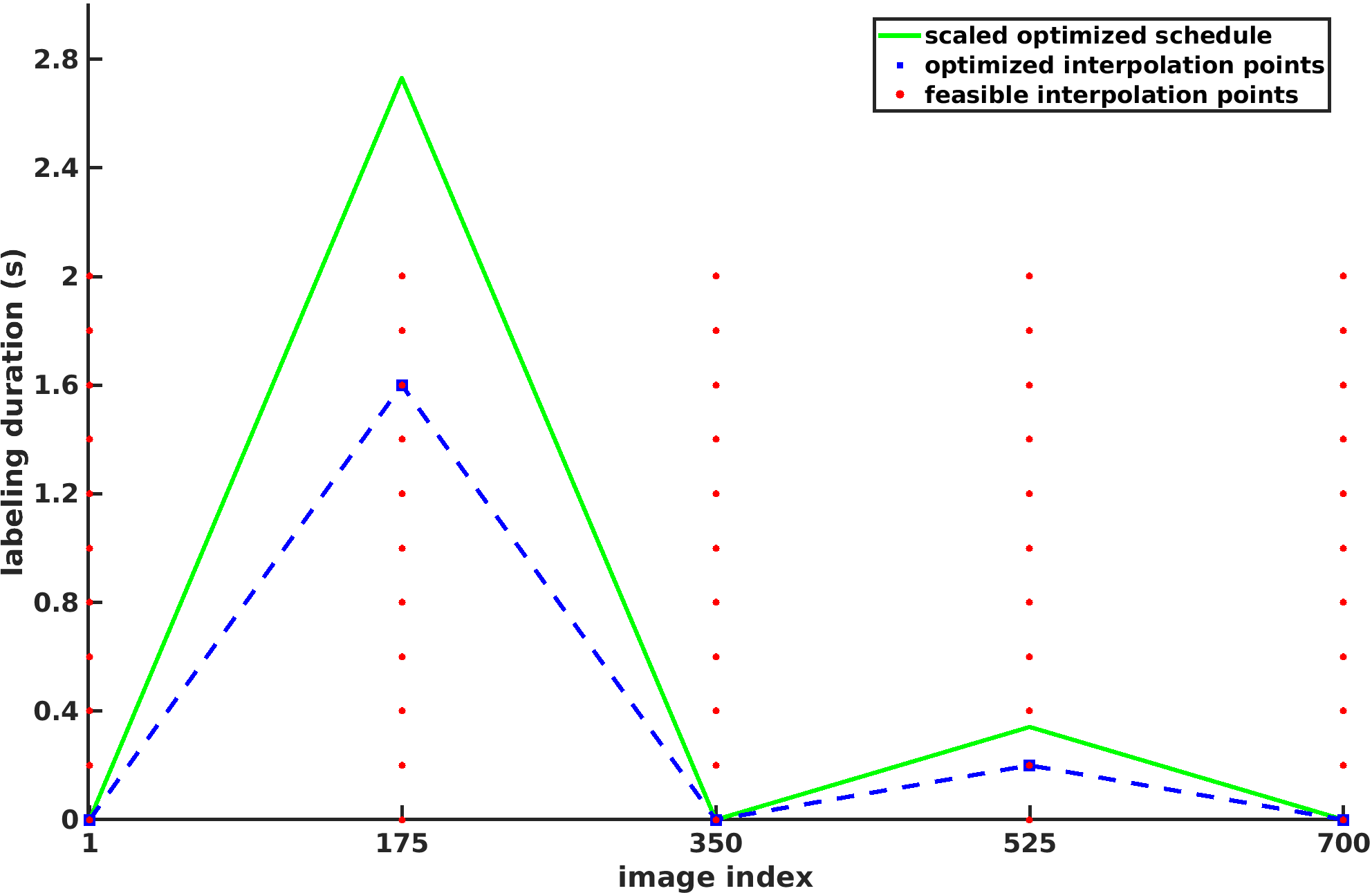}
\caption{The red dots depict the feasible points for interpolating between or exploring the labeling space. The blue squares depict the label durations for the five interpolation points that, once scaled, leads to our optimized schedule. The green line depicts this scaled, `optimal' (among other candidates) schedule.}
\label{fig:optSched}
\end{center}
\vspace{-0.1in}
\end{figure}

The set of feasible candidate labeling schedules, $\mathscr{V}$, consists of ASL timing sequences with variable labeling times but fixed pre and post labeling delays. TR was allowed to vary depending on the labeling duration. The labeling durations were described using a linear interpolation of 5 points in the `labeling space', each a fixed number of frames apart, as shown in \fref{fig:optSched}. We choose linear interpolation as it allows us to explore this labeling space effectively. It also allows for flexible finer sampling of the labeling space, albeit at the expense of optimization time, upon increasing the number of interpolation points or possible labeling durations. Feasible schedules are scaled to be a fixed total duration before the cost function is evaluated. Because there is a trade-off between sacrificing scan duration and sacrificing precision, the total duration may vary based upon the required precision. For our work, the scan duration is set to $600$s, and the number of acquired images are 700. \fref{fig:optSched} also depicts the optimized scan. Having obtained an optimized labeling schedule, we further minimize the predicted precision of flow estimates by trying several pseudo-random label-control-silence schedules while maintaining that the number of each are equal in our schedules. In section \ref{resOpt}, we compare the theoretical performance of this labeling schedule to two suboptimal ones commonly encountered in MRF literature, with $|\Theta| = 250$.  While the neural network-based estimation framework described in 
section \ref{nnetT} is not unbiased, these numbers serve as a good indicator for the performance of a schedule in terms of precision in estimates obtained from it.

\subsection{Estimation with Neural Networks}\label{nnetT}
\begin{figure}[H]
\begin{center}
\includegraphics[height=2.0in]{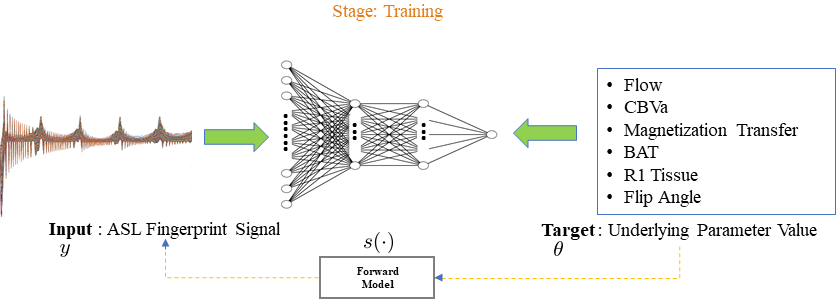}
\caption{Diagram depicts a neural network regressor used as an estimator in our work, in the training stage. The targets for the training and the inputs are related through the forward model depicted in \fref{fig:model} with additive noise. Separate networks are trained for the different unknowns in the model. Once trained, the estimators can predict the generating parameters for a new fingerprint.}
\label{fig:nNetEstimator}
\end{center}
\vspace{-0.1in}
\end{figure}

We use a neural network based framework to estimate 5 hemodynamic properties of relevance in our model. Namely, these are the Perfusion $f$ (CBF), the Bolus Arrival Time (BAT) or $\delta$, the Cerebral Blood volume in artery ($CBV_a$), Magnetization Transfer Rate ($K$), and the tissue relaxation time, henceforth called $T_{1}$ for simplicity. Additionally, we also estimate a field map of the Flip angles enacted by the scanner. Separate neural networks are used in the estimation of each parameter. The reason for moving away from the combined neural network framework used in our previous work \cite{Lahiri2018OptimizedRegression} is to avoid the need for the relative weighting of targets during network training. \tref{nnetTab} provides the architectural specifications of the networks used . Each neural network was constructed to have the least amount of hyperparameters necessary to provide consistent losses in training and test data.
\begin{center}
\begin{table}[h!]
\centering
\begin{tabular}{ |c||c|c|c|c| } 
 \hline
 Parameter & Depth & Architecture (nodes per layer) & min value & max value \\ 
 \hline\hline
 Perfusion & 3 & 10-10-10 & 0 mL/100g/min & 90 mL/100g/min\\ 
 \hline
 CBVa & 3 & 10-10-10 & 0  & 0.015 \\ 
 \hline
 BAT & 2 & 10-5 & 0.3 s  & 3.0 s\\ 
 \hline
 MTR & 4 & 10-10-5-5 & 0 s$^{-1}$ & 0.03 s$^{-1}$\\ 
 \hline
 $T_1$ & 1 & 20 & 0.33 s & 3.33 s\\
 \hline
 Flip & 1 & 20 & 48\degree & 112\degree\\ 
 \hline
\end{tabular}
\caption{Description of the neural network architectures used in estimating hemodynamic parameters in our signal model, as well as the respective maximum and minimum values of the ranges used in the training data. The `Architecture' column provides the number of nodes in every layer, separated by hyphens, starting from the input. Each node in the network learns a weight and a bias during training. The input to the networks are fingerprints generated from our designed optimized sequence, which has 700 frames.}
\label{nnetTab}
\end{table}
\end{center}
To train our networks, we use $6\times10^6$ samples of synthetic fingerprints generated from the model described in section \ref{modT}, with added real white Gaussian noise with variance $0.01$, along with the corresponding generating parameters (\fref{fig:nNetEstimator}). The same training dataset was used across all neural networks. We used an independent uniform prior on values of each parameter for generating this data. The associated parameter ranges are also depicted in \tref{nnetTab}. Our choice of using independent uniform priors across all ground truth parameters in the training data was motivated by being able to perform well at estimating possible anomalies in combinations of hemodynamic properties. (For example, elevated arterial transit time, but normative perfusion etc.)  Each signal was normalized by the value of the first frame in the fingerprint. This is also enacted upon signals obtained from the scanner, thereby ensuring consistency during testing and training. The cost function used to train the neural networks is Mean Square Error, the optimizer associated was ADAM \cite{Kingma2014Adam:Optimization}, and the non-linearities were implemented as ReLU-s. Training times for the networks were roughly $15-20$ mins. Once trained, the network was tested on a gamut of test datasets described in section \ref{Res}.

We noisy simulated data from the signal model instead of real data for training for two reasons: (i) ground truth estimates for real data are difficult to obtain, and in a wide-scale multiparametric setting such as ours, would suffer from granularity owing to the use of dictionary based methods in calculating the ground truth for training, or are simply biased because they were obtained from non-MRF techniques (ii) limited availability of real training data would pose a significant risk of overfitting neural networks (especially deeper ones).

\subsection{Signal Preconditioning}\label{preconT}
Presence of scanner drift or cardio noise and breathing can cause severe distortions in the fingerprints from the hypothesized model \cite{Lund2006Non-whiteImpact,Glover2000Image-BasedRETROICOR}. However, the labeling scheme modulates the perfusion information into the high frequency bands of the fingerprint signal similar to \cite{Liu2005AMRI,Mumford2006EstimationFMRI}. This property, combined with the fact that the aforementioned nuisances generally manifest as low frequency components, motivated us to high-pass filter the fingerprints both during training and testing of the neural networks associated with perfusion, bolus arrival time, magnetization transfer rate and the cerebral blood volume fraction. We applied a $4$-th order Butterworth filter with a cutoff at $0.05$ Hz (when assuming the fingerprint was sampled at $1$ Hz) for this purpose.
\subsection{Data Collection}

\subsubsection{Simulated Anthropomorphic Pathological Phantoms}\label{AntPhantSec}
\begin{figure}[h!]
\begin{center}
\includegraphics[height=4.0in]{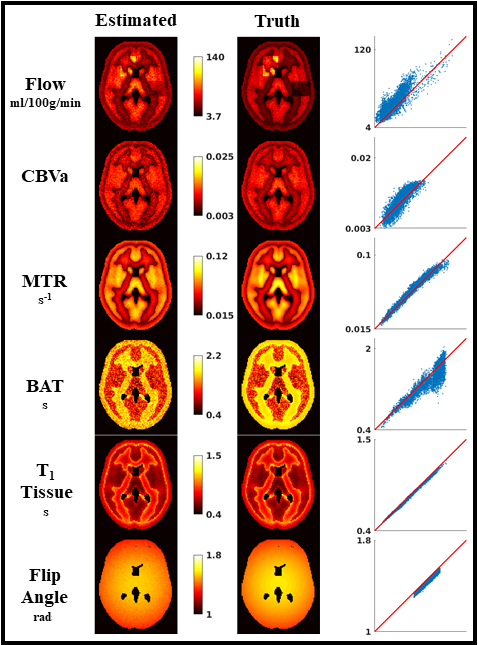}
\caption{Performance of proposed neural network based estimation on the simulated dataset described in section \ref{AntPhantSec}. The first column depicts the predictions from the networks, while the second shows the corresponding ground truth parameter images. The third column are 'truth-vs-predicted' scatter plots of the former columns.}
\label{fig:AntPhant}
\end{center}
\vspace{-0.1in}
\end{figure}

We synthesized a set of test data from hemodynamic parameter maps that closely reflects the corresponding spatial distributions in a digital phantom generated from standard gray and white matter maps (SPM12) \cite{Friston2007StatisticalImages}. We then introduced regions of abnormally elevated and reduced perfusion to it to quantify the performance of our methods on a range of normative and pathological parameter values. Real AWGN with standard deviation $0.01$ was added to the fingerprints after generation. When estimating perfusion, CBVa, and  BATs, the data was high-pass filtered as explained in section \ref{preconT}. \fref{fig:AntPhant} compares the predictions with the corresponding ground truths.

\subsubsection{In vivo experiments}\label{invivoexpt}
In-vivo data was acquired on a 3T General Electric MR750 scanner. The imaging parameters were: a single slice placed above the ventricles, single shot spiral readout, nominal resolution = 3.5$\times$3.5$\times$7 mm$^3$, matrix size = $64\times64$, bandwidth = 125 Hz and TE = 5ms, FOV = 240 mm. We tested our methods on data acquired from six human subjects. For four of these subjects, we also acquired data with two unoptimized MRF-ASL scan designs and compared our optimized scan design to them using similar estimation techniques (same neural network architecture and training target distribution in \tref{nnetTab}). Our goal was to show the benefits of labeling sequence optimization in ASL fingerprinting. For `suboptimal scan 1', we sampled the labeling durations from a uniform random distribution, while in `suboptimal scheme 2', the durations decrease linearly with the image index. Both schemes were designed to be 700 frames and 600s long. The metric for this comparison was the normalized standard deviation of parameter estimates obtained from the numerical CRLB evaluation described in section \ref{optT}. 

For all six subjects, we also performed a 409s multi-PLD ASL \cite{Woods2019AExperiments,Alsop1996ReducedFlow} experiment with 40 PLDs, involving a single average over label-control pairs at each PLD. The post-labeling delays were chosen according to the protocol presented in \cite{Woods2019AExperiments}. The CBF, CBVa, BAT and $T_1$ maps obtained from these were compared to those from our methods.

We used a two-stage estimation technique to generate quantitative parameter maps from the multi-PLD data. In the first stage, we estimated the tissue $T_1$ and M0 maps at every voxel by applying a least squares fit using the model in eqn. \ref{modeleq}. For this stage, all other parameters in the equation were fixed to nominal, or where applicable, normative values. Next, the entire process was repeated  for estimating the CBF, CBVa and BAT at each voxel, but using M0 and $T_1$ values obtained in the previous stage. The MTR and flip angles were held fixed throughout.

\section{Results}\label{Res}

\subsection{Optimized Scan Design}\label{resOpt} We compared the predicted performance of our optimized scan design against two suboptimal ASL MRF scan designs described in section \ref{invivoexpt}. \tref{compSubOptTab} lists the predicted normalized standard deviation in estimates of each parameter for all three labeling schemes, and the total weighted design cost associated with each scheme. 
 \begin{center}
\begin{table}[H]
\centering
\begin{tabular}{ |c||c|c|c| } 
 \hline
 Parameter & optimized scheme & sub opt. scheme 1 & sub opt. scheme 2 \\ 
 \hline\hline
 Perfusion & 46.4 & 51.0 & 46.1 \\
 \hline
 CBVa & 17.2 & 21.9 & 19.5 \\ 
 \hline
 BAT & 1.3 & 2.1 & 1.6 \\ 
 \hline
 MTR & 121.1 & 124.2 & 137.1 \\ 
 \hline
 $T_1$ & 0.6 & 0.4 & 1.1 \\ 
 \hline
 Flip & 0.4 & 0.2 & 0.8 \\ 
 \hline\hline
 \textbf{Cost} & \textbf{233.4} & \textbf{250.8} & \textbf{252.3} \\
 \hline
 
\end{tabular}
\caption{Predicted normalized standard deviation of parameter estimates (in \%) for ASL-MRF labeling schedules used in our comparisons. The last row shows the overall weighted design cost associated with each scheme based on eqn. \eqref{objfun}.}
\label{compSubOptTab}
\end{table}
\end{center}
\subsection{Simulated Anthropomorphic Pathological Phantoms}
\fref{fig:AntPhant} depicts the estimated maps from the Anthropomorphic Pathological phantom simulation, and the corresponding ground truth parameter images in the first two columns. From these images, we also generated `truth vs predicted' scatter plots for each estimated parameter map to better visualize the accuracy and precision of our methods, shown in the third column of \fref{fig:AntPhant}.
For a more quantitative evaluation of the performance of our methods, Table \ref{CorrTab} shows the correlations between the voxel values of the truth and estimated parameter maps across multiple scan designs. 
\begin{center}
\begin{table}[H]
\centering
\begin{tabular}{ |c||c|c|c| } 
 \hline
 \multirow{2}{*}{Parameter} &\multicolumn{3}{c|}{ Correlation w/ truth (\%)} \\ \cline{2-4}
 & opt. scan & sub opt. scan 1 & sub opt. scan 2 \\
 \hline
 Perfusion & $\bm{86.7}$ & 71.9 & 68.2\\ 
 \hline
 CBVa & $\bm{86.7}$ & 80.0 & 81.6 \\ 
 \hline
 MTR & $\bm{98.4}$ & 95.4 & 0\\ 
 \hline
 BAT & $\bm{91.7}$ & 90.0 & 89.1 \\ 
 \hline
 $T_1$ & 98.6 & $\bm{99.6}$ & 98.0\\ 
 \hline
 Flip & $\bm{99.7}$ & 98.6 & 84.7\\ 
 \hline
\end{tabular}
\caption{Correlation of each estimated parameter map with the corresponding ground truth map in the anthropomorphic digital phantom.}
\label{CorrTab}
\end{table}
\end{center}

\begin{figure}[H]
\begin{center}
\includegraphics[height=4.0in]{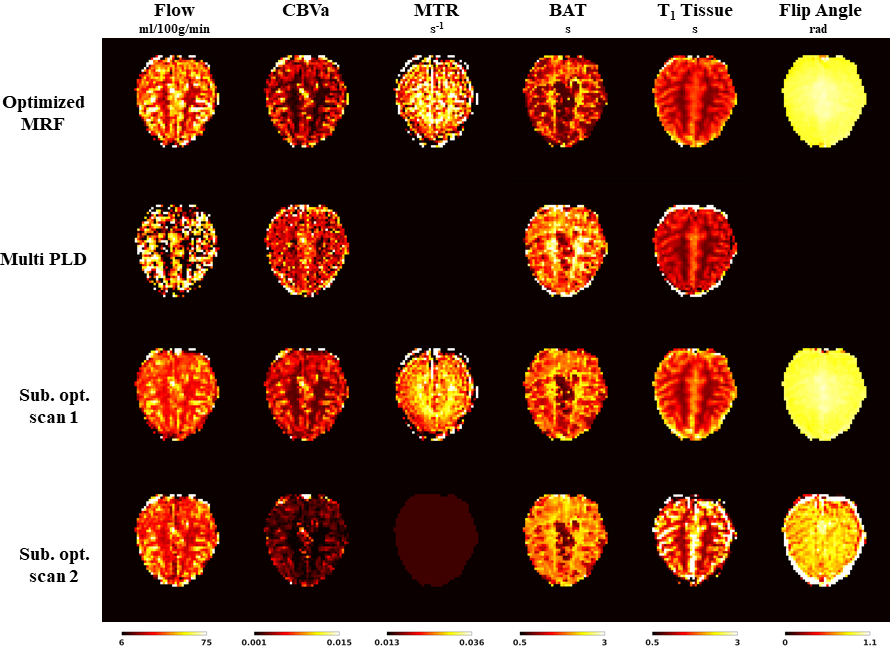}
\caption{Comparison of the parameter estimates from various tested methods for a single subject. None of the methods used any spatial smoothing of the estimated maps or the ASL signal volume.}
\label{fig:HumComp}
\end{center}
\vspace{-0.1in}
\end{figure}

\subsection{In vivo performance} 
\begin{figure}[h!]
\begin{center}
\begin{tabular}{cc}
\includegraphics[height=2.0in]{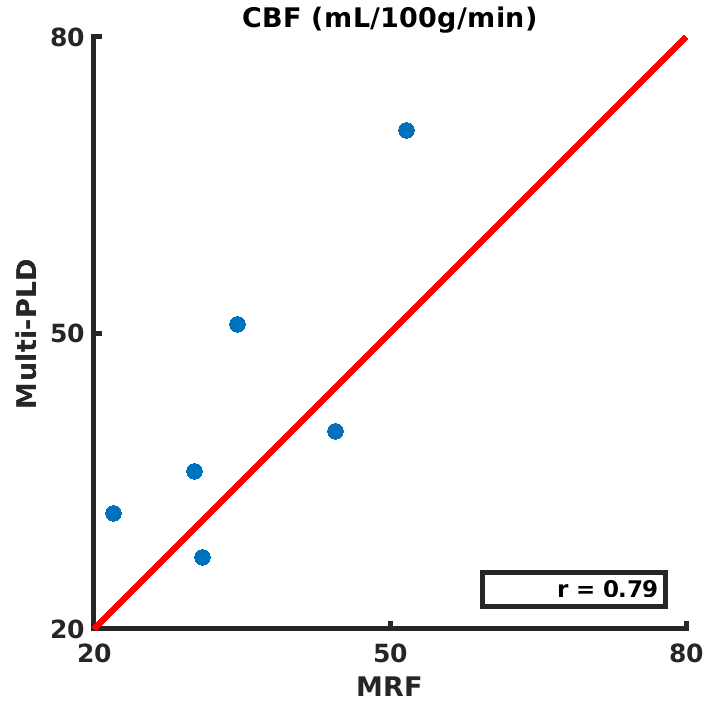}     & \includegraphics[height=2.0in]{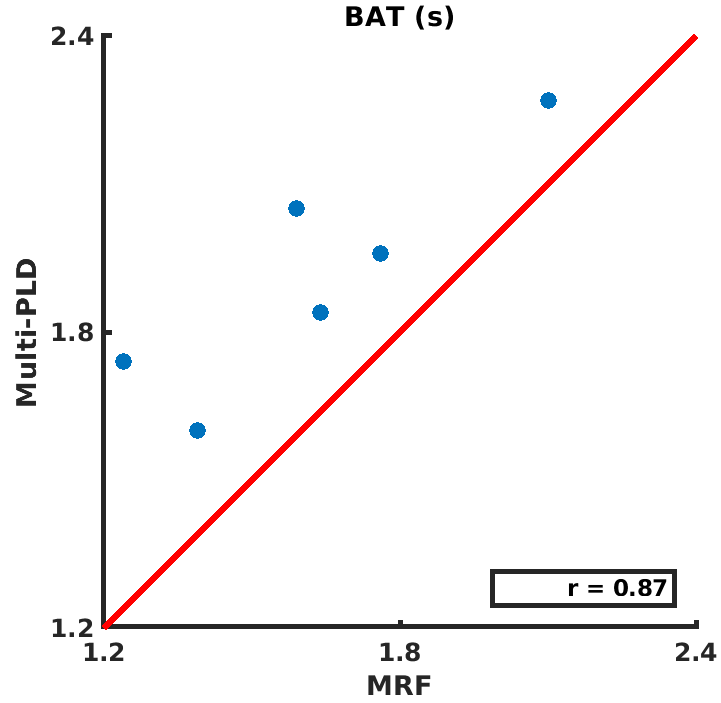}  \\
\hspace{0.1in}(a)\hspace{0.1in} & \hspace{0.1in}(b)\hspace{0.1in}\\
\includegraphics[height=2.0in]{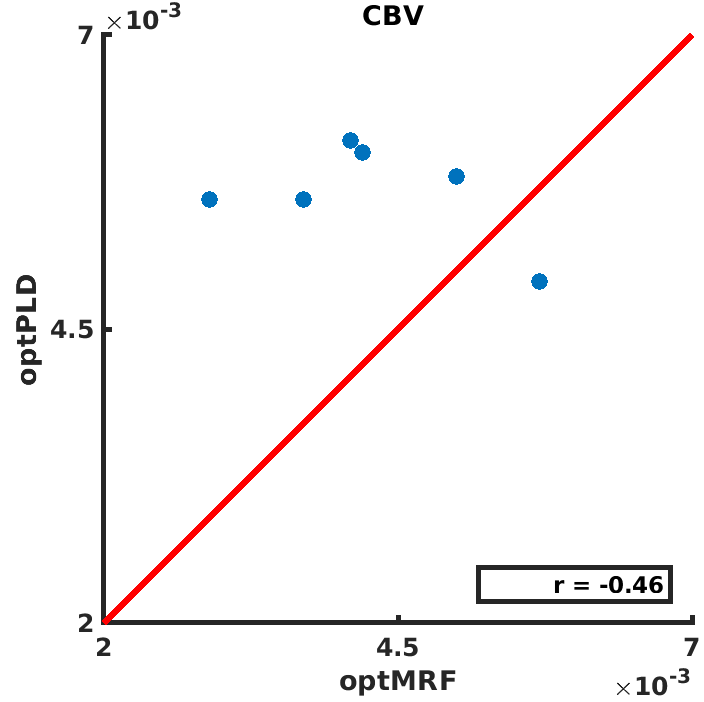}     & \includegraphics[height=2.0in]{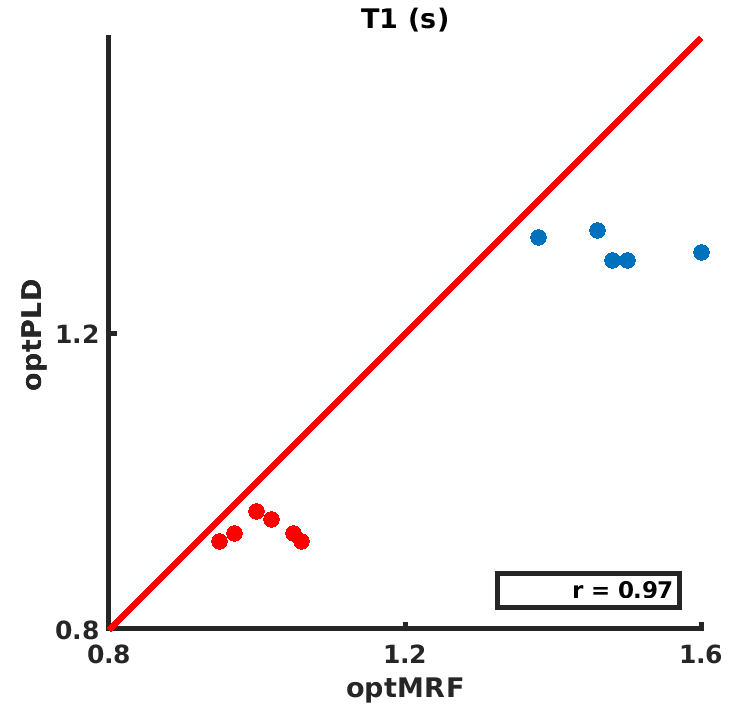}  \\
\hspace{0.1in}(c)\hspace{0.1in} & \hspace{0.1in}(d)\hspace{0.1in}
\end{tabular}
\caption{Scatter plots of slice-wide average estimates from optimized MRF vs multi-PLD of: (a) gray matter CBF, (b) gray matter BAT, (c) gray matter CBVa and (d) gray (blue dots) and white (red dots) matter $T_1$.}
\label{fig:scatterMRFPLD}
\end{center}
\vspace{-0.1in}
\end{figure}

It took approximately 1s to estimate each $64\times64$ parameter map for the MRF methods using the designated neural networks for the task on a 12GB NVIDIA Titan X Pascal GPU. The two stage fit for the multi-PLD data required approximately 1200s to estimate four $64\times64$ maps on a Intel Xeon E5-2650 with 40 cores. \fref{fig:HumComp} compares the six estimated maps from a single human subject across all evaluated techniques. 
To gauge a sense of agreement between estimates from regression based ASL Fingerprinting and multi-PLD methods, \fref{fig:scatterMRFPLD} compares the mean gray matter CBF, BAT and CBVa across the six subjects, as well as the associated average $T_1$s for both gray and white matter using scatter plots. 
\fref{fig:roiResidueComp} shows a `goodness-of-fit' comparison between the acquired signal and a `synthetic' signal produced from the modeled equations (\ref{modeleq})-(\ref{modeleqfin}) averaged over a region of interest. The synthetic signals were obtained by passing the parameters estimated from the neural networks through the ASL signal model.



\section{Discussion}\label{Dis}
This work established a CRLB based optimization method for labeling durations for improving the information within an MRF-ASL scan, as a means to get more precise estimates from it for a fixed scan time. This enables us to `get the most' out of available scan time, and is of particular importance because of the trade-off between total scan time and precision of estimates, regardless of the estimator. Of course, it would be possible to reduce the scanning duration at the expense of overall precision. We also adopted a neural network regression based estimation framework to avoid the granularity/imprecision of dictionary-search based estimators for problems with many parameters like ASL-MRF. The methods provided estimates for six parameters in both gray and white matter regions in the brain, and we  validated our methods in silico using a simulated anthropomorphic phantom, and in vivo against a multi-PLD method as well as other suboptimal ASL-MRF scans. In both cases, the CRLB predictions were reflected in the performance of various methods in a relative sense. The following subsections elaborate on our observations from section \ref{Res}.
\subsection{Optimized Scan Design}
From the predicted standard deviations in \tref{compSubOptTab}, it is apparent that the optimized scan either at least performs comparably, or outperforms the other two at precisely estimating all relevant parameters. In particular, the overall cost function for the optimized labeling schedule is significantly lower than that for the others. This hints its potential for improved precision at jointly estimating all the modeled dependencies in the ASL signal. The MTR parameter contributes significantly to the overall variance of estimates, but incorporating it into our model may provide additional information about tissue health and reduce bias in estimates of perfusion. We also find that while a lot of variation in the labeling durations can help in the estimation of $T_1$ and Flip angles, it can be detrimental when attempting to estimate Perfusion (CBF), Blood Volume Fraction (CBVa) or Arterial Transit Times (BAT).
\subsection{Simulated Anthropomorphic Pathological Phantoms}
\fref{fig:AntPhant} illustrates that there is good agreement between the estimated and the ground truth maps across all parameters. Additionally, our estimation is able to capture both the abnormally elevated and diminished regions of flow in the perfusion maps. This property is a consequence of: (i) optimizing the scan design over a large parameter range, (ii) training the neural network
using a wide range of training parameters (\tref{nnetTab}).

\tref{CorrTab} shows that for most estimated parameters, when compared to the suboptimal scans, our optimization leads to comparable or better correlation coefficients between truth and estimates. For perfusion in particular, we noted that the optimized scan yielded estimates that were significantly more aligned to the truth than the other scans. It was also intruguing to note that `suboptimal scan 2' regressed the same value of MTR ($\approx$ 0.015) for all inputs, thereby returning a correlation coefficient of 0\%. Our conjecture is that this may be due to the neural network being unable to learn from the training data due to the insensitivity of the fingerprints to MTR. We also observed that the correlation coefficent associated with perfusion for this specific scan was lower than the others, even though its predicted normalized standard deviation was comparable to that of the optimized scan. We therefore hypothesize that predictions of low normalized standard deviation may not always translate to high correlation between truth and estimates in the case of a biased estimator. This is because, variable estimation bias at different points in the parameter space may lead to low correlation, despite the precision in estimates. 

\subsection{In vivo performance}
\fref{fig:HumComp} illustrates that the performance of our designed method is relatively consistent with the predictions from the Cramer-Rao Bound (see \tref{compSubOptTab}). The map corresponding to the magnetization transfer rate looks the noisiest, while parameters like $T_1$ or BAT look much cleaner. The distinction between gray and white matter regions is also apparent across all relevant maps, even without any spatial smoothing or SNR boosting methods.

The comparisons between the optimized and suboptimal MRF scans were also in accordance with our expectations based on \tref{CorrTab}. As evident in the depicted subject, `suboptimal scan 2' fails to estimate the MTRs, and the estimated $T_1$ maps have unreasonably high values in gray matter. Moreover, even the flip angle map for this scan shows significant artefactual contrast between gray and white matter, which are absent in the other MRF methods. The maps yielded by `suboptimal scan 1' show agreement with the optimized scan, but we observe that the MTR maps from the former are less informative, and exhibit more artifacts. These trends between the optimized and suboptimal scans were noted to be consistent across all four subjects studied.

\fref{fig:HumComp}, shows that the MRF methods are able to estimate CBF, BAT and CBVa values in white matter, while the multi-PLD method fails to do so: instead, the corresponding maps show near-zero perfusion in white matter regions, along with abnormally high transit times. The obtained CBVa map from the latter method also appears extremely noisy, and is unable to pick out vasculature in the slice as effectively. We hypothesize that the poor performance in the multi-PLD method was due to the fact that its acquisition parameters were optimized for BAT and CBF, without regard for CBVa.

In \fref{fig:scatterMRFPLD} (a)-(b), we found the estimates from optimized MRF-ASL and multi-PLD to be fairly consistent, as observed from the high correlation coefficients between the two methods, for measurements of CBF as well as BAT. However, we note that the multi-PLD method consistently reports higher BATs in gray matter. \fref{fig:scatterMRFPLD} (c) shows a similar scatter plot for CBVa, but there is little agreement between the methods. This may be due to the fact that the estimates of CBVa from the multi-PLD method have a lot of variance across the brain, for reasons explained above. The $T_1$ estimates for the two methods are compared in \fref{fig:scatterMRFPLD} (d), with the red dots indicating white matter $T_1$s, and the blue ones indicating gray matter $T_1$s. We see that while the optimized MRF method generally yields higher values of $T_1$, the measurements agree well.

\begin{figure}[h!]
\begin{center}
\includegraphics[height=1.75in]{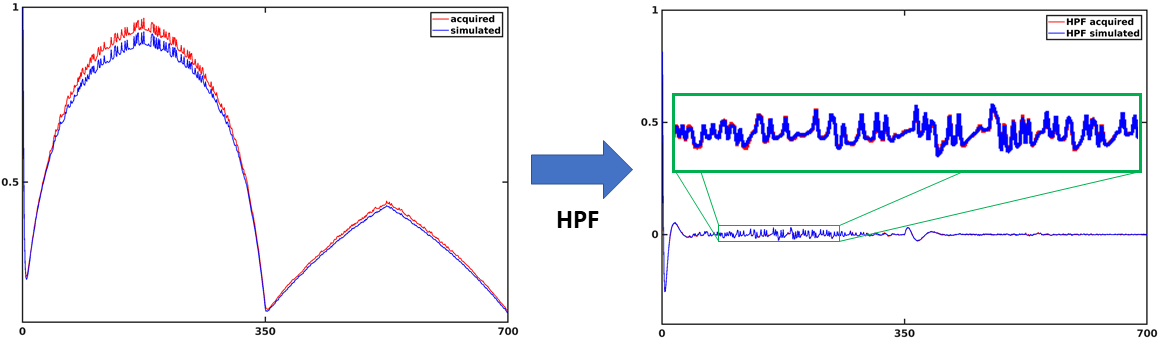}
\caption{ROI-averaged comparison of acquired signals and simulated signals which were generated from our model, based on neural network estimates. The left and right panes show the signals before and after high pass filtering, respctively.}
\label{fig:roiResidueComp}
\end{center}
\vspace{-0.1in}
\end{figure}

\fref{fig:roiResidueComp} reinforces that high pass filtering the fingerprint signals significantly improves the signal fidelity in the small, high frequency components, that correspond to manifestations of hemodynamic phenomenon. This increased agreement is because using a high pass filter removes low frequency components related to receiver drift, etc. that introduces discrepancies that were not accounted for in the model.
\section{Conclusion}\label{Conc}
In conclusion, we have developed a framework for optimizing the scan design for MRF-ASL that yields more precise estimates in gray and white matter, than suboptimal scan designs for MRF-ASL, and a reference multi-PLD method. We also introduced a neural network regressor for fast precise estimates from ASL fingerprints. These methods were able to estimate six parameters from a single 600s ASL scan in a very short processing time, and significantly improve upon the state-of-the-art results in MR Fingerprinting ASL.
\section{Acknowledgements}
NIH R21EB021562
\bibliographystyle{plain}
\bibliography{main}
\end{document}